\documentclass[prl,aps,twocolumn,showpacs,preprintnumbers,floats,amsmath,amssymb]{revtex4}

\usepackage{graphicx}
\usepackage{dcolumn}
\usepackage{bm}

\def \be{\begin{equation}}
\def \ee{\end{equation}}

\begin{document}

\title{Variational wave functions, ground state and their overlap}

\author{Christophe Mora$^{1,2}$}
\author{Xavier Waintal$^3$}
\affiliation{$^1$~Laboratoire Pierre Aigrain, D\'epartement de Physique,
Ecole Normale Sup\'erieure,
24 rue Lhomond, F-75005 Paris, France}
\affiliation{$^2$~Institut f\"ur Theoretische Physik,
Heinrich-Heine Universit\"at, D-40225 D\"usseldorf}
\affiliation{$^3$~Nanoelectronics group, Service de Physique de l'\'Etat Condens{\'e},
CEA Saclay, 91191 Gif-sur-Yvette cedex, France}

\date{\today}

\begin{abstract} An intrinsic measure of the quality of a variational wave function is given by its overlap with the
ground state of the system. We derive a general formula to compute this overlap when quantum dynamics in imaginary time is 
accessible. The overlap is simply related to the area under the $E(\tau)$ curve, i.e.
the energy as a function of imaginary time. This has important applications to, for example, quantum Monte-Carlo
algorithms where the overlap becomes as a simple byproduct of routine simulations.
As a result, we find that the practical definition of a good variational wave function for quantum 
Monte-Carlo simulations, {\it i.e.}  fast convergence to the ground state, is equivalent to a good
overlap with the actual ground state of the system.
\end{abstract}

\maketitle

Variational wave functions (VWF) are very valuable tools to study interacting quantum systems.
Examples are numerous and can be found in many fields of physics, like Helium 4~\cite{clark2006}, 
high Tc superconductors~\cite{anderson1987} or the fractional quantum Hall effect~\cite{laughlin1983}. 
Given a variational wave function $\Psi_V$, it is not difficult to sample  $|\Psi_V|^2$
from which one obtains the expectation value of many physical observables. It remains 
difficult however to determine to which degree $\Psi_V$ is a good approximation of the
true ground state $\Psi_0$ of the system. A first answer to this question lies in the
variational theorem itself: as the variational energy is always higher than the ground
state energy, one attempts to get the VWF with the lowest energy. The drawback of this criterion
is that VWFs with similar variational energies can convey very different physics as the
energy is very sensitive to the short range part of the VWF, {\it i.e.} when two particles are 
close to each other,
but rather weakly to the long range part. This is  unfortunate as other physical 
observables may depend crucially on the VWF. Another (standard) possibility is to pick the VWF that has 
the lowest variance of its energy. The variance criterion has some advantages over the
energy~\cite{foulkes2001} as it is a more absolute criteria: a zero variance means that the
VWF is an eigenstate of the system (but not necessarily the ground state).

In this letter, we concentrate on a more intrinsic criterion, namely maximizing the overlap $\Omega$:
\be 
\Omega \equiv \frac{|\langle\Psi_V|\Psi_0\rangle|^2}
{\langle\Psi_0|\Psi_0\rangle \langle\Psi_V|\Psi_V\rangle}
\ee
of $\Psi_V$ with the actual ground state $\Psi_0$ of the system.
A direct calculation of $\Omega$ would require the complete knowledge of the 
ground state $\Psi_0$, which is extremely computationally demanding. The main result of this letter,
namely Eqs. \eqref{def} and \eqref{kappa}, is that $\Omega$ can be related to the energy, upon
projection of the initial VWF in imaginary time. Hence, $\Omega$ can be simply obtained  as the byproduct
of quantum Monte-Carlo simulations. In  this letter, we focus on this particular method to determine $\Omega$.
We would like to emphasize, however, that 
 our main result together with inequalities \eqref{ineq}, \eqref{ineq2} and \eqref{ineq3} are universal and can
be applied to other methods.

The development of zero temperature quantum Monte-Carlo (QMC) techniques (Diffusive or 
Green function Monte-Carlo) relies on the availability of good VWFs. Although those techniques
allow one to access ground state properties, VWFs play a crucial role for two reasons.
 Firstly, 
QMC uses a VWF to properly sample the Hilbert space through ``importance sampling''.
An homogeneous sampling of Hilbert space would mean spending a very large amount of time sampling regions
of the Hilbert space that have almost no contribution to $\Psi_0$. Secondly, even though 
QMC simulations can give  interesting information on the system they do not necessarily explain the
physical mechanisms involved. Valuable insight
can often be obtained just by looking at how the VWF is constructed. For instance the short and long range 
behavior of a Jastrow type VWF, or more dramatically the analytical structure of the Laughlin wave function
in the fractional quantum Hall effect conveys in itself useful information. With QMC comes an empirical 
criterion for what is a good VWF: it should allow the simulation to converge from $\Psi_V$ to $\Psi_0$  
as fast as possible. We shall see that this criterion actually coincides with maximizing the 
  overlap $\Omega$ of $\Psi_V$ with the actual ground state $\Psi_0$.
\begin{figure}
\vglue +0.05cm
\includegraphics[width=6cm]{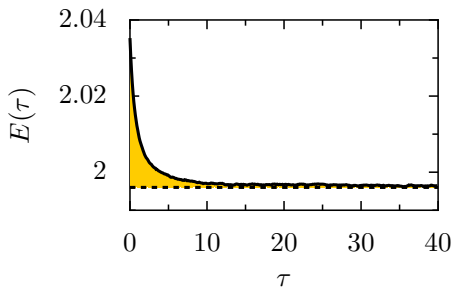}
\caption{\label{Edetau} (color online) Typical curve for the energy $E(\tau)$ as a function of 
the imaginary time $\tau$. The thick line is a typical
trace (with high precision) while the dashed line corresponds to the asymptotic value $E_0$. The logarithm of the overlap $\Omega$ is  
simply given by the shaded area between the two curves (see text). The data have been obtained for the model
of dipolar bosons, defined Eq.\eqref{hamilto2}, with very high precision.}
\end{figure}

The idea to maximize the overlap as a variational criterion was  put forward twenty years 
ago~\cite{reatto1982,masserini1984,masserini1987}. It was
shown that for a certain class of VWF, merely of the Jastrow type and its extensions, maximizing $\Omega$
amounts to choosing the Jastrow function that  reproduces exactly the (mixed estimator of the) static structure
function. Hence the practical criterion was to optimize the static structure function obtained from variational Monte-Carlo 
by comparison with that
coming out of QMC simulations. Here we choose a different route and show that $\Omega$ can be computed directly as a simple byproduct 
of a QMC run. In QMC techniques one starts with an initial wave-function $\Psi_V$ and then integrates in a stochastic way
the imaginary time Schr\"odinger equation
$\partial_\tau \Psi = - H \Psi$, the formal solution being
$\Psi(\tau)=e^{-\tau H} \Psi_V$. For a large $\tau$,  $\Psi(\tau)$
actually converges toward $\Psi_0$. In practical simulations, one usually computes the energy 
\be
E(\tau)=\frac{\langle\Psi_V|H|\Psi(\tau)\rangle}{\langle\Psi_V|\Psi(\tau)\rangle}
=\frac{\langle\Psi_V|H e^{-\tau H}|\Psi_V\rangle}{\langle\Psi_V|e^{-\tau H}|\Psi_V\rangle}
\ee
until $E(\tau)$ has converged to its asymptotic value, the exact ground state energy $E_0$. An example of
a typical $E(\tau)$ curve is shown in Fig.~\ref{Edetau} for the system discussed at the end of this letter. By introducing $\kappa$ as
\be
\label{def}
\Omega \equiv e^{-\kappa},
\ee
the principal result of this letter is that $\kappa$ is simply given by the shaded area in Fig.~\ref{Edetau}.
Or more precisely,
\be
\label{kappa}
\kappa=\int_0^\infty d\tau\ \  [E(\tau)-E_0]
\ee
It is therefore straightforward to obtain $\kappa$ from a QMC simulation. 

Eq.(\ref{kappa}) calls for a few comments.
(i) as mentioned above, Eq.(\ref{kappa}) combines a theoretical criterion, maximizing the overlap
between $\Psi_V$ and $\Psi_0$, and a practical criterion that is fast convergence to $E_0$. Conversely, a wave-function
that has a low variational energy but converges very slowly toward $E_0$ has a poor overlap with the actual ground state.
(ii) $\kappa$ is not a substitute for the variational energy (or variance) as it requires  a full QMC simulation.
(iii) In contrast with energy and energy variance, $\Omega$ is dimensionless and hence has an absolute meaning. For example
$\Omega=0.97$ means that the VWF captures $97\%$ of the 
ground state. 
(iv) In some instances, one is interested in the thermodynamic $N\rightarrow\infty$ limit, where $N$ is
 the number of particles in the system. 
It is easy to be convinced that the generic behavior of $\Omega$ is an exponential decrease with
 $N$. 
This can be seen  for instance in a non-interacting Bose-Einstein condensate, 
where any error in the one particle wave-function appears to the power 
$N$ in the many-body wave function. More generally the energy is
usually an extensive quantity and hence  $\kappa$ scales linearly with $N$ for large $N$. In that case, the correct
measure of the accuracy of a VWF is $\kappa/N$ so that $\Omega = (0.99)^N$ for example means 
that the VWF captures ``per particle'' $99\%$ of the ground state. $\kappa/N$ usually shows weak finite size effect upon increasing $N$. 
(v) In practice, we have found that the accuracy of $\kappa$ is similar to that obtained
for other physical quantities like density or density-density correlations but slightly less precise than that achieved
for $E_0$ (Relative precisions of $10^{-4}$ or better are routinely obtained for the latter).

{\it Proof of Eq.(\ref{kappa}).} The proof is straightforward. We introduce the pseudo-partition function
$Z(\tau)\equiv\langle\Psi_V|\Psi(\tau)\rangle=\langle\Psi_V|e^{-\tau H}|\Psi_V\rangle$, which, in analogy to the finite
temperature partition function is related to the energy through $E(\tau)=-\partial_\tau \log Z(\tau)$.
Defining $\bar\Psi_0$ as the ground state normalized to unity, $\Psi(\tau)$ converges  towards
$\Psi(\tau)\rightarrow \sqrt{Z(2\tau)}\bar\Psi_0$ for large $\tau$. Using the definition of
$\Omega$, and $Z(\tau)$ one obtains
\be
\label{toto}
\Omega=\lim_{\tau\rightarrow\infty} \frac{[Z(\tau)]^2}{Z(0)Z(2\tau)}.
\ee
Further, from $E(\tau)=-\partial_\tau \log Z(\tau)$ one finds
$\log (Z(\tau)/Z(0))=-\int_0^\tau E(\tau)d\tau$ and $\log(Z(2\tau)/Z(\tau))=-\int_\tau^{2\tau} E(\tau)d\tau$.
In the latter, for $\tau$ large enough, $E(\tau)\sim  E_0$ such that $Z(2\tau)/Z(\tau)\sim \exp(-E_0 \tau)$.
Collecting terms together in Eq.(\ref{toto}), we arrive at Eq.(\ref{kappa}).

{\it Link with mixed estimators.} One drawback of QMC calculations is that when the quantum average of an observable $\hat A$ is
measured, the mixed estimate
$A_{\rm MX}=\langle\Psi_V|\hat A|\Psi_0\rangle/\langle\Psi_V|\Psi_0\rangle$ naturally emerges. For some observables
it is possible to actually calculate the correct quantity $A_0=\langle\Psi_0|\hat A|\Psi_0\rangle/\langle\Psi_0|\Psi_0\rangle$
 using, for instance, forward walking techniques~\cite{buonaura1998}.  However this concerns mainly local quantities and does not apply to non-local ones.
For the latter, one has to rely on the extrapolation formula, $A_0\approx 2 A_{\rm MX}-A_V$ where
$A_V=\langle\Psi_V|\hat A|\Psi_V\rangle/\langle\Psi_V|\Psi_V\rangle$. This formula~\cite{barnett1991} is only valid when $\Psi_0$ is close
to $\Psi_V$, otherwise it is uncontrolled.
If $\hat A$ is definite positive, it takes the form 
 $\hat A=d^\dagger d$ and the overlap allows one to obtain the following lower bound for $A_0$,
\be
\label{ineq}
A_0\ge \frac{[A_{\rm MX}]^2}{A_{\rm V}}\ \Omega
\ee
A particularly interesting example is the condensate fraction of a Bose condensate where the 
operator $d^\dagger$ creates a particle in the $k=0$ state.
Unfortunately, this lower bound is only useful for small systems like atoms or molecules, or for extremely good VWFs, since the exponentially decreasing 
nature of the overlap quickly makes it meaningless. To prove Eq.(\ref{ineq}), one simply writes Schwartz inequality, 
$|\langle C|B\rangle|^2\le\langle C|C\rangle\langle B|B\rangle$ with $|C\rangle=d|\Psi_0\rangle$
and $|B\rangle=d|\Psi_V\rangle$

{\it Link with the energy gap.} Another application of the calculation of the
overlap is that it gives access to an upper bound value for the gap of the system,
i.e. to information on the excitation spectrum. Indeed, it is straightforward
to show that the difference $\Delta$ between the first excited state and $E_0$
obeys:\cite{eckert1930}
\be\label{ineq2}
\Delta\le\frac{E_V -E_0}{1-\Omega}
\ee
Again, such an upper bound value is usually useless in the thermodynamic limit ($N\rightarrow\infty$) where
there usually is a continuum of excitations. It can be used however for molecular or atomic systems.
In this case, one should look for a wave-function  close to the first excited state. Such a VWF has
a rather low energy but also a low overlap with the ground state. In practice the VWF, being close 
to an eigenstate, has a low variance, and as the variance is $-\partial E/\partial\tau|_{\tau=0}$, the $E(\tau)$ curve converges 
slowly to the ground state energy, hence resulting in a small overlap. The quality of this upper bound value depends on the projection 
of the VWF on the excited states other than the first one and the inequality becomes an equality when the 
VWF has projections only on the ground state and first excited state.

\begin{figure}
\vglue +0.05cm
\includegraphics[width=8cm]{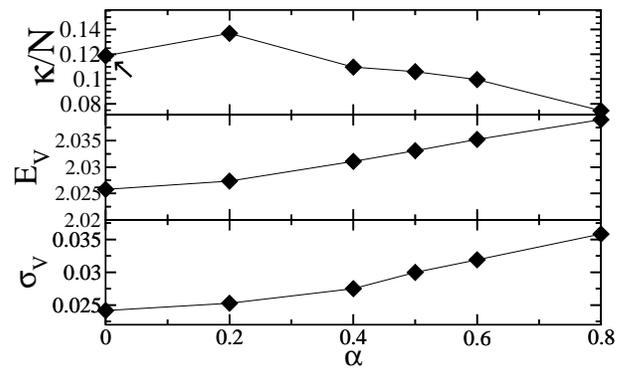}
\caption{\label{fig:over} Upper panel: parameter $\kappa/N$ measuring the overlap between the guiding function and the actual ground state of the system as a function of the symmetry breaking parameter $\alpha$. The arrow indicates that the point at $\kappa(\alpha=0)$ is lower than the correct value, a small $\alpha\ne 0$ being needed for the simulation to 
fully converge to the ground state. Middle panel: variational energy $E_V$ as a function of $\alpha$. Lower panel:  square root of the quantum variance $\sigma^2_V$ of the VWF as a function of the symmetry breaking parameter $\alpha$. All data are taken for $32$ bosons in a  $32\times 56$ grid at $r_S=28.6$.}
\end{figure}

{\it Application.} To illustrate the utility of the above discussion, we now turn to a specific example where the 
calculation of the overlap can be particularly useful. The system discussed below is close to a first order liquid-solid transition. On one
hand the chosen VWF is close to the liquid state so that optimizing the variance or the energy leads closer to the
liquid state. On the other hand the true ground state is the crystal as indicated by the calculation of the overlap.  
More precisely, we consider a system
of $N$ bosons in two dimensions with a repulsive dipolar interaction. This system is subject to current research as
it is a candidate for a realizing a crystal in ultracold atom experiments~\cite{astrakharchik2007,buchler2007}. 
A detailed study of the model will
be presented elsewhere~\cite{mora2007b}, and we focus here on the relevance of $\Omega$. The scaled 
Hamiltonian takes the form,
\be
\label{hamilto2}
H = - \frac{1}{r_s} \sum_{i=1}^{N} \nabla_i^2 + 2 \sum_{i<j}
\frac{1}{|{\bf r}_i - {\bf r}_j|^3}.
\ee
where $r_S$ controls the relative strength of the dipolar interaction over the kinetic one.
This model shows a first order transition at $r_S=r_S^*\approx 27$: For $r_S\le r_S^*$ the system is in a Bose-Einstein phase
while for $r_S\ge r_S^*$ the system crystallizes into a triangular lattice. 
Note that this crystal-Bose Einstein transition has been discussed recently in 
Refs.~\cite{astrakharchik2007,buchler2007}.
We use a VWF of Bijl-Jastrow form: 
\be\Psi_{V} ({\bf r}_1,\ldots,{\bf r}_N) = 
\prod_{i=1}^N \Phi_1({\bf r}_i) \prod_{j<k} \Phi_2 (|{\bf r}_j - {\bf r}_k|)
\ee
where the Jastrow part includes two-body correlations,
$\Phi_2 (r) = \exp \left( -2 \sqrt{\frac{r_s}{r}} e^{-r/A} \right)$. 
The one-body part allows to break translational symmetry and interpolates from
a condensate-like VWF to a crystal-like VWF.
Noting $\Delta y$, the distance between lattice sites of the crystal along the $y$ axis, we define the 
vector ${\bf q}_1 = (0,2 \pi / \Delta y)$ in the reciprocal lattice. Vectors ${\bf q}_{2/3}$
are obtained by rotating ${\bf q}_1$ by an angle of $2 \pi /3$ and $-2\pi/3$.
We choose a one-body trial function
\be
\Phi_1 ( {\bf r} ) = \prod_{i=1,2,3} \left ( 1+ \alpha \cos ( {\bf q}_i \cdot {\bf r} ) \right),
\ee
whose maxima reproduce the triangular lattice expected for the crystal.
This function is well-suited to describe quantum melting since it interpolates between a flat
liquid(BEC)-type pattern for $\alpha=0$ to a triangular crystal form for $\alpha \ne0$.
Setting $r_S=28.6$, i.e. slightly above the crystalline transition, we investigate our
VWF as a function of the translational symmetry breaking parameter $\alpha$. The results are presented in Fig.~\ref{fig:over}.
The details of our algorithm can be found in ref~\cite{mora2007b,waintal2006}. 
In particular we used the Green function Monte-Carlo technique on a spatial grid with a filling factor
$\nu = 1/56$ particles per site. 

The system is clearly in a crystal state, as demonstrated  by 
computing the static structure factor, and hence a good variational ansatz is expected for $\alpha \ne 0$.
Fig.~\ref{fig:over} shows however that 
both the variational energy and variance indicate a minimum for $\alpha=0$. On the other hand $\kappa$
decreases with $\alpha$ indicating that the phase with $\alpha\ne 0$ is the correct one. This is an
extreme case where the traditional criterion on the energy and variance actually gives a false answer even 
qualitatively while the overlap indicates the correct answer.

{\it Conclusion.} It is interesting to note that both the standard criteria used to characterize 
a VWF (variational energy and variance), as well as the overlap discussed in the present letter, are
all different aspects of the energy-imaginary time $E(\tau)$ curve: the variational energy 
$E_V=E(\tau=0)$, the variational variance $\sigma_V^2=-\partial E/\partial\tau|_{\tau=0}$ 
and the overlap is the integral of $E(\tau)$. Hence the different criteria give informations 
on different characteristics of $E(\tau)$. While no general statement can be made, in many 
instances the VWFs used are a fair description of the ground state. In these cases $E(\tau)$ 
looks typically like the curve shown in Fig.\ref{Edetau} and, in particular, has a positive 
curvature. Assuming a positive curvature, we find that the various criteria are related to each other
as
\be\label{ineq3}
\Omega\le \exp \left[-(E_V -E_0)^2/(2\sigma^2_V)\right],
\ee 
 obtained using the fact that $E(\tau)$ lies above its 
tangent at $\tau=0$. Although the above inequality is not completely general,
it gives a rough estimate of the overlap in many practical cases. One can note that
strictly speaking, the above estimate decreases when the variance decreases. It is therefore imperative
 that the variance and variational energy are optimized simultaneously. 

 To conclude this letter, we have shown that from the data given by usual QMC calculations, it is possible
to extract relevant information on the quality of the variational wave-function used. 
We stress that our main result, Eq.~\eqref{kappa}, applies generally to any projection technique in imaginary time
and not only to QMC.

\acknowledgments
We thank O. Parcollet for fruitful discussions and S. Dhilon for many useful comments.
C.M also acknowledges support from the DFG-SFB Transregio 12.


\newcommand{{{\PRB}}}{{{Phys. Rev. B}}}\newcommand{{{\PRA}}}{{{Phys. Rev. A}}}\newcommand{{{\PRL}}}{{{Phys. Rev. Lett}}}\newcommand{{{\NPB}}}{{{Nucl. Phys.}}}\newcommand{{{\RMP}}}{{{Rev. Mod. Phys.}}}\newcommand{{{\ADV}}}{{{Adv. Phys.}}}\newcommand{{{\EPJB}}}{{{Eur. Phys. J. B}}}\newcommand{{{\EPJD}}}{{{Eur. Phys. J. D}}}

\end{document}